# Predicting protein folding dynamics using sequence information

*Ezequiel A. Galpern, Federico Caamaño and Diego U. Ferreiro*

**Authors**

Ezequiel A. Galpern, Federico Caamaño and Diego U. Ferreiro*:* Laboratorio de Fisiología de Proteínas, Facultad de Ciencias Exactas y Naturales, Universidad de Buenos Aires and Consejo Nacional de Investigaciones Científicas y Técnicas, Instituto de Química Biológica de la Facultad de Ciencias Exactas y Naturales (IQUIBICEN-CONICET), Buenos Aires CP1428, Argentina

**Abstract**

Natural protein sequences somehow encode the structural forms that these molecules adopt. Recent developments in structure-prediction are agnostic to the mechanisms by which proteins fold and represent them as static objects. However, the amino acid sequences also encode information about how the folding process can happen, and how variations in the sequences impact on the populations of the distinct structural forms that proteins acquire. Here we present a method to infer protein folding dynamics based only on sequence information. For this, we will rely first on the obtention of a precise 'evolutionary field' from the observed variations in the sequences of homologous proteins. We then show how to map the energetics to a coarse-grained folding model where the protein is treated as a string of foldons that interact. We then describe how, for any given protein sequence of a family, the equilibrium folding curve can be computed and how the emergence of protein folding sub-domains can be identified. We finally present protocols to analyze how mutations perturb both the folding stability and the cooperativity, that represent predictions for a deep-mutational scan of a protein of interest.

A google colab implementation of the method is available:

https://colab.research.google.com/github/eagalpern/folding-ising-globular/blob/master/notebooks/custom_potts_simulation_colab.ipynb

**Keywords**

folding pathway, folding mechanism, Ising, evolution, DCA, Potts,

**Introduction**

Natural protein molecules are truly amazing objects. Over the last century it has become clear that these molecules perform their various biological tasks by folding the polypeptide chains into dynamic ensembles of three dimensional structures. Today, protein folding is understood in the conceptual framework of the Energy Landscape Theory, which posits that evolved proteins are minimally frustrated heteropolymers which navigate physical landscapes that have the overall form of a rough funnel, for which there is a fundamental correlation between structure and energy [1]. Precise, fast and robust folding is not a general property of most amino-acid chains, but results from the selection of specific sequences that, when folded, minimize the energetic conflicts between its parts [2]. The amazing variety of structures that natural protein molecules display is somehow coded in the linear strings of amino acids [3], [4]. Statistical analysis of natural sequence information must then allow us to reverse-engineer the basic aspects of the folding codes [5]. Recently, massive machine learning schemes have been proven successful to predict a three-dimensional structure for a given sequence input [6], [7], apparently solving the structure-prediction problem. However, these black-boxes tell us nothing about the mechanisms by which the actual polypeptides acquire the structural forms, how broad the structural ensembles are or how sequence variations impact the folding dynamics.

From a biological standpoint, protein folding appears to be a major driving force in protein sequence evolution [8], [9], [10]. As such, the patterns observed when aligning multiple sequences of evolutionary related proteins must somehow account for the energetics of amino-acid interactions. Mutations can be interpreted as *de facto* perturbations of the folding landscape that have passed the sieve of evolution. This accounts for the theoretical underpinning of the now widely used amino-acid substitution matrices, domain recognition programs and protein phylogeny reconstructions [11], [12], [13]. Moreover, recent developments of statistical models to account for natural sequence variations have been proved useful for analyzing the energetic effect of single point mutants and pair-mutations couplings [14], [15], [16].

It was discovered when studying the folding of single-domain proteins that, provided there are strong enough native interactions within them to compensate for their entropy loss, distinct protein sections may fold at different periods almost independently. Panchenko et al. have dubbed these units "foldons" as they can fold in a single coordinated step [17]. The exon-foldon correlation has been supported by recent evidence that conserved exons exhibit a strong independent foldability [18]. Thus it may be possible to use this primary structure partitioning to identify common foldons for each family and examine their folding dynamics, drawing on the annotation of intron-exon borders in several genomes.

We present here a simple model to predict folding dynamics by leveraging the sequence information learned by inverse statistical evolutionary models and a coarse-grained description of protein structure. We summarize how by assuming that protein stability has been the most relevant evolutionary constraint (see Note 1), it is possible to infer folding curves, free energy profiles and subdomain emergence for any given protein sequence.

**Materials**

The essential components for an Ising coarse-grained folding simulation include:

1. A target protein sequence.
2. A Multiple sequence alignment of proteins homologous to the target sequence.
3. A partition of the sequence into contiguous folding elements, often referred to as *foldons*.

We propose using a sequence-based energy function that evaluates the probability of observing a protein sequence of a given length within an equilibrated ensemble. These "evolutionary" energy parameters can be computationally derived from a Multiple Sequence Alignment (MSA) using methods such as Direct Coupling Analysis (DCA) [19], [20].

1. Target Protein Sequence

A target amino acid sequence is required to initiate the folding simulation. This sequence can represent the full length of a protein, or a fragment (e.g., a domain) that is assumed to fold independently of the rest of the chain. Variations in folding dynamics are expected across protein families, particularly for protein topologies rich in alpha-helices (see Note 3). Notably, the impact of all possible single-site mutations can be estimated directly from the wildtype simulation without requiring additional simulations [21].

2. Multiple Sequence Alignment (MSA)

To derive a sequence-based evolutionary energy, a set of homologous and aligned sequences is necessary. A Multiple Sequence Alignment (MSA) can be obtained from the Pfam database [22] (now hosted by InterPro [23]) by searching for the protein family containing the target sequence. Alternatively, an MSA can be constructed using the UniParc database [24], where homologous sequences are identified and aligned to the target sequence using tools like *jackhmmer* [25]. Both approaches rely on scoring single-site similarities and penalizing gaps, parameters that one must be careful to tune for obtaining a meaningful MSA. For Pfam alignments, to ensure the MSA aligns with the target sequence, only positions present in the target sequence should be retained. The quality of the evolutionary field to be obtained relies on the depth, coverage and representativeness of the MSA used for the learning. In other words, a large enough, diverse and clean MSA is required. We describe some relevant considerations to fulfill these requirements. Protein sequences present in databases represent actual proteins (or predicted sequences from genomic analysis) that the community has put together in databases for diverse reasons, and as such are biased by the organisms studied and the phylogenetic relationships among them. To minimize phylogenetic bias, MSA sequences can be globally analysed in order to detect overall similarity and, assuming that they come from closely related organisms, downgrade the relative contribution. This can be done clustering the complete, unaligned sequences with CD-hit [26] at a 90% cutoff. This cutoff is not stringent but has been proven sufficient for large datasets. A faster alternative is to cluster the aligned sequences using as similarity the normalized Hamming distance between them and applying the same cutoff. A weight defined as *1/n* is assigned to each sequence i, where $n_i$ is the number of sequences in the *i*-th cluster, so the sum of clusters is the effective number of sequences. All statistical analyses should be performed using these sequence weights. The likelihood of accurately learning the Potts model parameters increases with the alignment depth, the number of effective sequences of the MSA or simply the sum of sequence weights. For reliable inference of Potts model parameters, it is generally recommended to have an effective number of sequences that is at least 5 to 10 times the sequence length *L* [27]. For instance, a protein of length *L = 100* would require at least 500-1000 effective sequences, while *L = 200* would require 1000-2000. We highlight that many sequences in raw alignments often do not cover the N-terminus or C-terminus regions (and sometimes in between regions), harming the accuracy of prediction that a model can reach. Certain positions present a very high fraction of gaps, increasing the noise-to-signal ratio during the Potts model learning, being an usual practice to remove from the MSA sequences that have a frequency of gaps higher than a given threshold, i.e. 25%.

3. Folding Elements (Foldons)

The protein sequence must be divided into distinct, non-overlapping folding elements, or foldons, which are groups of amino acids that fold and unfold cooperatively. While a common foldon assignment for each protein family (or MSA) is recommended, the method can accommodate sequence-specific partitions. We describe here some heuristics to assign foldons, according to the protein architecture and the hypothesis to test.

- Repeat Proteins:
  For periodic or repeat proteins, foldons can be defined as individual repeats or segments of repeats. The periodic nature of these proteins simplifies foldon assignment, and the folding dynamics can be studied by treating each repeat as a single foldon. Also, a periodic partition of the protein can be obtained by separating the repeats into multiple folding units [28], [29] or grouping them into larger foldons.

- Secondary structure elements:
  For non-symmetric globular proteins, secondary structure elements (e.g., alpha-helices, beta-strands) can serve as foldons. This approach leverages the natural structural organization of the protein to define folding units.

- Exon-Based Partitioning:

    Highly conserved exon boundaries within a protein family can be used to define foldons, as there is often a correspondence between exons and folding units [18].

- Arbitrary Segments:

    Foldons can also be defined based on known structural segments or arbitrary partitions for testing specific hypotheses. This flexibility allows researchers to explore the impact of different folding unit definitions on the simulation results.

- Neutral Model:

    Alternatively, foldons can be assigned using a neutral model, where segment sizes are sampled from a geometric distribution. Multiple neutral partitions can be generated, and results can be averaged across them to account for variability in foldon definitions. This approach is useful for exploring the robustness of the results to different partitioning schemes.

**Methods**

The general procedure for predicting protein folding dynamics using sequence information consists of 6 steps (Box 1)

1. Learn the Potts model parameters from a MSA.
2. Estimate the Selection Temperature.
3. Assign the coarse-grained folding hamiltonian.
4. Set up the Monte Carlo parameters and run simulations.
5. Visualize and analyze results for a single protein.
6. Extra: Exploit family results for protein design.

Model definition and methods pipeline are summarized in figure 1.

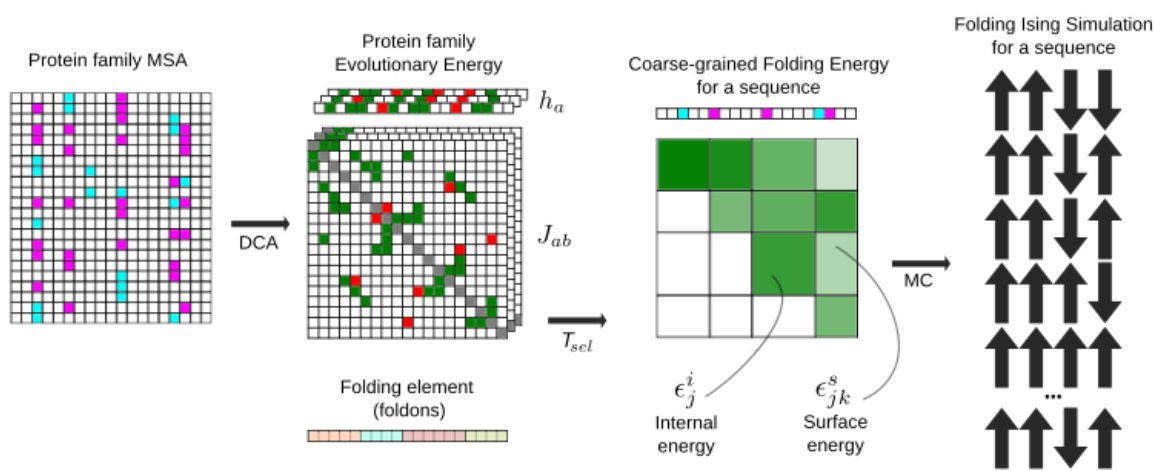

*Figure 1. Model Definition. We learn the Evolutionary Energy field parameters from a Multiple Sequence Alignment (MSA) using a Direct Coupling Analysis (DCA). Given a set of folding units, or foldons and the Selection Temperature of the family, we extract the coarse-grained folding energy for each sequence to input a finite-chain Ising model. The folding dynamics of the sequence is computed using a Monte Carlo simulation.*

1. Learn the Potts model parameters from a MSA.

The evolutionary energy parameters can be inferred from a MSA through a *Direct Couple Analysis* (DCA). A Potts model is defined, containing local fields and pairwise couplings, to account for the occurrence of a given amino acid in a given position and for the co-occurence of amino acids in pairs of positions. The energy function of the model for a given sequence σ is

$$E(\sigma) = -\sum_{a=1}^{L} h_a(\sigma_a) - \sum_{a<b}^{L} J_{a,b}(\sigma_a, \sigma_b),$$

where the parameters to learn are the local fields $h_a$ and the pairwise couplings $J_{ab}$. The exact inference of the model parameters requires the calculation of the partition function, which is computationally intractable. Several approximation strategies have been developed, giving rise to many DCA implementations, including Boltzmann Machine Learning (bmDCA) [30], [31], PseudoLikelihood Maximization (plmDCA) [32] and Mean Field approximation (mfDCA) [19]. A comprehensive review can be consulted in [33]. For first-time DCA users, we recommend starting with plmDCA, implemented in the *pydca* library available for Python. However, we highlight that the folding model may benefit from using more accurate learning methods as bmDCA (see note 2 for other alternatives). The *pydca* library requires a MSA in fasta format and it computes sequence weights via a similarity clustering as explained in *Materials,* using a cutoff defined by the user. Also, the learning uses two regularization parameters for avoiding overfitting. One of these parameters is the local fields regularization term, and it was observed that the outputs were robust to it, so a default value for this parameter of 1 can be used [32]. The other parameter, which is the pairwise coupling regularization term, seems to be more critical, therefore requiring a scan of values to determine a precise choice of this parameter. Based on previous observations, a default value of 20 was set for this parameter and used throughout experiments. The accuracy of the learned Potts Model can be estimated by looking at how many contacts of a native reference structure can be predicted as strong couplings in the Potts model. In addition, a simple validation step is recommended: to compute the total energy of the target sequence and ensure it is distinguishable from scrambled sequences. A more stringent test is to use the energetic field to obtain artificial sequences. This can be done with Monte Carlo simulations of sequences given the energetic field. One can then compare the single-site and pair-site amino acid frequencies of the simulated sequences versus a test set of natural sequences and their measured energies. The artificial sequences must be statistically indistinguishable from the natural ensemble of the MSA if the Potts field is perfectly inferred. Also, the mean energy of random sequences must be set to zero. This condition is automatically satisfied when the zero-sum gauge [33] is imposed, as most DCA algorithms do by default. Alternatively, a pre-trained Potts model can be used if it matches the target sequence's family. Minor alignment or size discrepancies can be corrected, but the sequence must belong to the family used to train the model. A validation using the described tests is also recommended for pre-trained models.

2. Estimate the Selection Temperature.

The Selection Temperature $T_{sel}$ is the apparent temperature at which sequences of a particular family were selected by nature and quantifies how strong the folding constraints have been during evolution [34]. If experimental folding free energy changes upon mutations *ΔΔG* are available for at least a protein of the studied family, $k_BT_{sel}$ can be explicitly calculated as the proportionality constant between *ΔΔG* and the corresponding changes scores as the energy difference of the wild-type and mutant according to Potts field. Alternatively, assuming that the standard deviation of *ΔΔG* is nearly constant irrespectively of protein families, $T_{sel}$ can be estimated in a relative scale to a known reference, using the ratio of the standard deviation of evolutionary energy changes by single mutations [35]. The reference should be a $T_{sel}$ obtained from *ΔΔG* experimental data*,* for instance for PDZ family, $T_{sel}$ will be estimated as

$$T_{sel}\ \sigma(\Delta E) = T_{sel}^{PDZ}\ \sigma(\Delta E^{PDZ}),$$

where σ (*ΔE*) denotes the standard deviation of Evolutionary Energy differences upon point mutations averaged over homologous sequences.

3. Assign the coarse-grained folding hamiltonian.

The protein is modeled as an array of interacting folding elements (foldons) that can be either folded (F) or unfolded (U), as 2-state spin variables. The system is represented as a finite-size Ising chain of N elements, where the energy of a coarse-grained configuration, the Hamiltonian, is given by the free energy of the corresponding ensemble of microstates

$$H = -\sum_{j=1}^{N} \left[ T\, s_j (1 - \delta_{j,F}) + \epsilon_j^i\, \delta_{j,F} \right] - \sum_{j=1}^{N-1} \sum_{k>j} \epsilon_{jk}^s\, \delta_{j,F}\, \delta_{k,F},$$

where $T$ is the temperature and $\delta_{j,F}$ is the Kroeneker symbol taking value one if element $j$ is folded (F) and zero otherwise. If the element $j$ is folded, it has a specific internal folding free energy (averaged over the solvent) $\epsilon_j^i$. If two elements $j$ and $k$ are both folded, we consider also a surface energy $\epsilon_{jk}^s$, describing a specific interaction between the two foldons. If the element $j$ is unfolded we set the energetic contributions to zero, but there is an explicit entropic contribution given by the entropy $s_j$ of the available spatial configurations of the foldon. Hence, within this model a protein can unfold as a result of an increase in temperature $T$.

The energetic ($\epsilon_j^i$, $\epsilon_{jk}^s$) parameters are functions of the amino acid sequence $\sigma$. We then map the sequence-based evolutionary energy residue–residue couplings $J_{ab}(\sigma_a, \sigma_b)$ and local fields $h_a(\sigma_a)$ to the specific coarse-grained folding free energy terms for each sequence,

$$\epsilon_j^i = \epsilon^i(\sigma_j) = k_B\, T_{sel} \left[ \sum_{a\,\epsilon\,j} h_a(\sigma_a) + \sum_{a,\,b\,\epsilon\,j} J_{ab}(\sigma_a, \sigma_b) \right],$$

$$\epsilon_{kj}^s = \epsilon^s(\sigma_j, \sigma_k) = k_B\, T_{sel} \left[ \sum_{a\,\epsilon\,j,\,b\,\epsilon\,k} J_{ab}(\sigma_a, \sigma_b) \right],$$

where $k_B$ is the Boltzmann constant and $T_{sel}$ is the Selection Temperature.

To compute the entropic terms ($s_j$), we recommend to approximate $s_j$ to be independent of amino acid identity and strictly additive and to use an average entropy per residue of $s = 5\, cal\, mol^{-1} K^{-1} res^{-1}$ that was empirically fitted [29]. Therefore $s_j = L_j\, s$, being $L_j$ the sequence length of the foldon. Alternatively, if experimental Temperature denaturation scans of the protein of interest (or a protein of the family) and mutants is known, the entropic term can be fitted by minimizing the difference between the simulated and the experimental data [29].

4. Set up the Monte Carlo parameters and run simulations

Once the Hamiltonian has been set for a sequence, the Metropolis Monte Carlo simulation of the Ising Model must be set up. To describe the folding dynamics, the simulation must be repeated for several temperatures $T$, such as for the lower $T$, the protein is completely folded and for the higher one is completely unfolded. The temperature range can be chosen manually or using the automatic range set that ensures the completely folding and unfolding condition. Simulation total time, transient time and equilibration time parameters may be sensitive to the Ising chain size (number of foldons per protein) which we assume to be constant for each family. An autocorrelation analysis is recommended to choose the optimal parameters for each $T$. As the system presents a critical temperature (at the folding temperature if it has two states) at which the Monte Carlo parameters could diverge, the simulation algorithm is organized in a two-round scheme. During the first round, the simulation is done at a list of given temperatures (typically distributed in a given range) and the critical temperatures are detected according to fluctuations. The simulation is repeated in a second round only at critical temperatures but using an equilibration and transient time multiplied by an extra parameter ('critical points factor'), that can be chosen according to the autocorrelation analysis.

5. Visualize and analyze results for a single protein.

As the temperature increases, the folding elements are expected to transition from the folded to the unfolded state. For each element, the algorithm outputs the simulated thermal unfolding curves and single-element folding temperatures $T_f^j$ (figure 2A). For visualization, the protein tertiary structure is colored according to element $T_f^j$ (figure 2B). Such structure, and its alignment with the MSA used should be provided. Our implementation simply requires the starting position of the sequence in the PDB file.

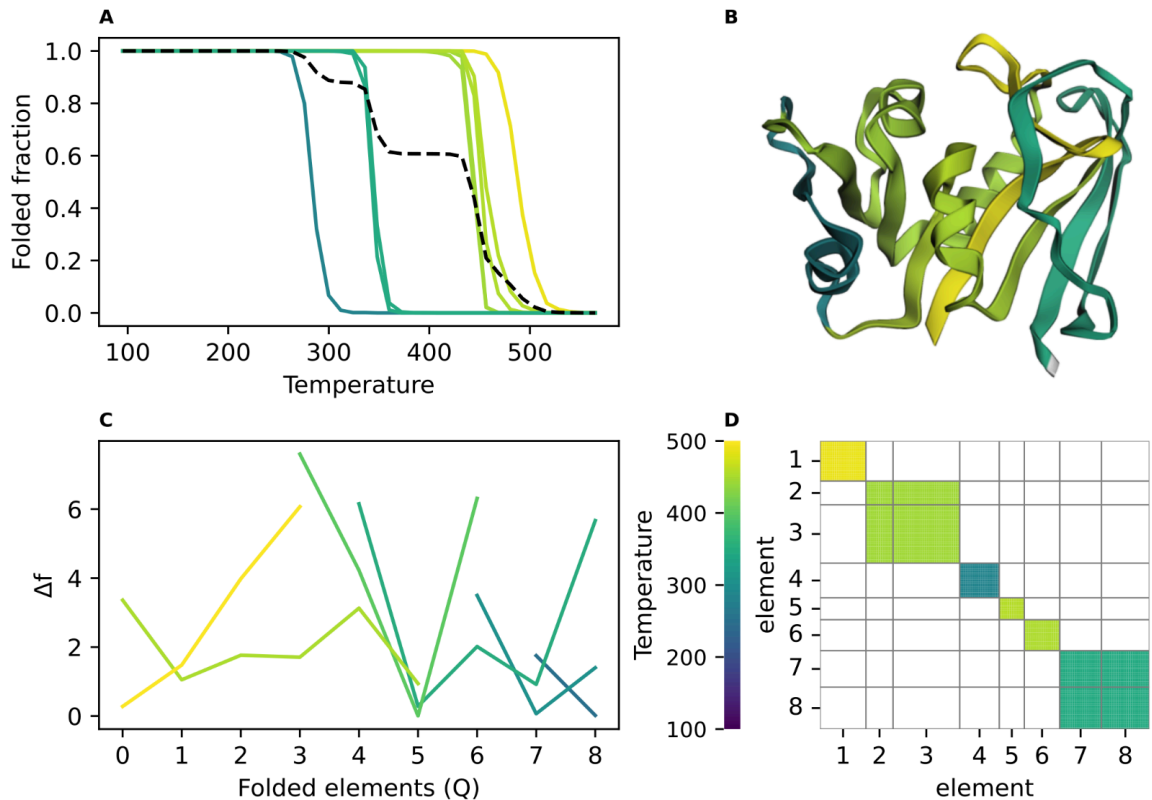

*Figure 2. Simulation results for a dihydrofolate reductase protein sequence (Uniprot ID A0A2D5XDZ3). A) Simulated thermal unfolding curves, for the complete protein (black dashed line) and for each element with solid lines, colors identify the folding temperature of each one (same as B, C and D). Yellow elements are the most stable ones. Protein folding temperature is $T_f$ = 411 K (B) The structure (PDB ID 7DFR) is colored according to the folding temperature of each element. (C) Free-energy profiles, colored by temperature, with the number of folded elements Q as reaction coordinate. Cooperativity Score is ρ = 5/7. (D) Apparent domain matrix and secondary structure colored by element folding temperatures.*

According to the energetics derived from the particular sequence, elements can unfold together at close temperatures. We define the emergence of an apparent subdomain between elements *j, k* if $|T_f^j - T_f^k| < 5$. Overlapping domains are separated into the minimum number of non-overlapping ones. If more than one separation is possible, temperature differences between domains are maximized. The subdomains are shown in a matrix plot (figure 2D).

For the complete protein, the fraction of elements folded at each temperature $m(T)$ is calculated taking all the foldons together. This curve can be contrasted to experimental thermal unfolding curves [29]. The protein folding temperature $T_f$ is estimated using a sigmoidal fit, approximating

$$m(T) = \frac{m_{max}}{1 + e^{a(T-T_f)}},$$

where $m_{max} \in [0, 1]$, taking out of the fitting isolated, extremely unstable elements.

A free energy profile is estimated from the probability of states $s$ with $Q$ folded elements with the Metropolis Monte-Carlo sampling (figure 2C). We considered together sampled states for simulations performed in a window of the 10 closest temperatures. The free energy profiles are calculated as

$$\Delta f(Q) = -k_B T \, log \left( \frac{\sum_{s|Q} N(s)}{\sum_s N(s)} \right),$$

where $T$ is the average temperature, $N(s)$ are the counts of state $s$ and $s|Q$ are the states with $Q$ folded elements.

As a measure of the folding cooperativity, we defined a Cooperativity Score $\rho = Q_{barrier} / (N - 1)$, the fraction of intermediary $Q$ that were not a minimum of $\Delta F(Q)$ for any $T$ in a protein with $N$ foldons. Together, the protein folding temperature, the cooperativity score and the domain emergence summarize the simulated folding dynamics.

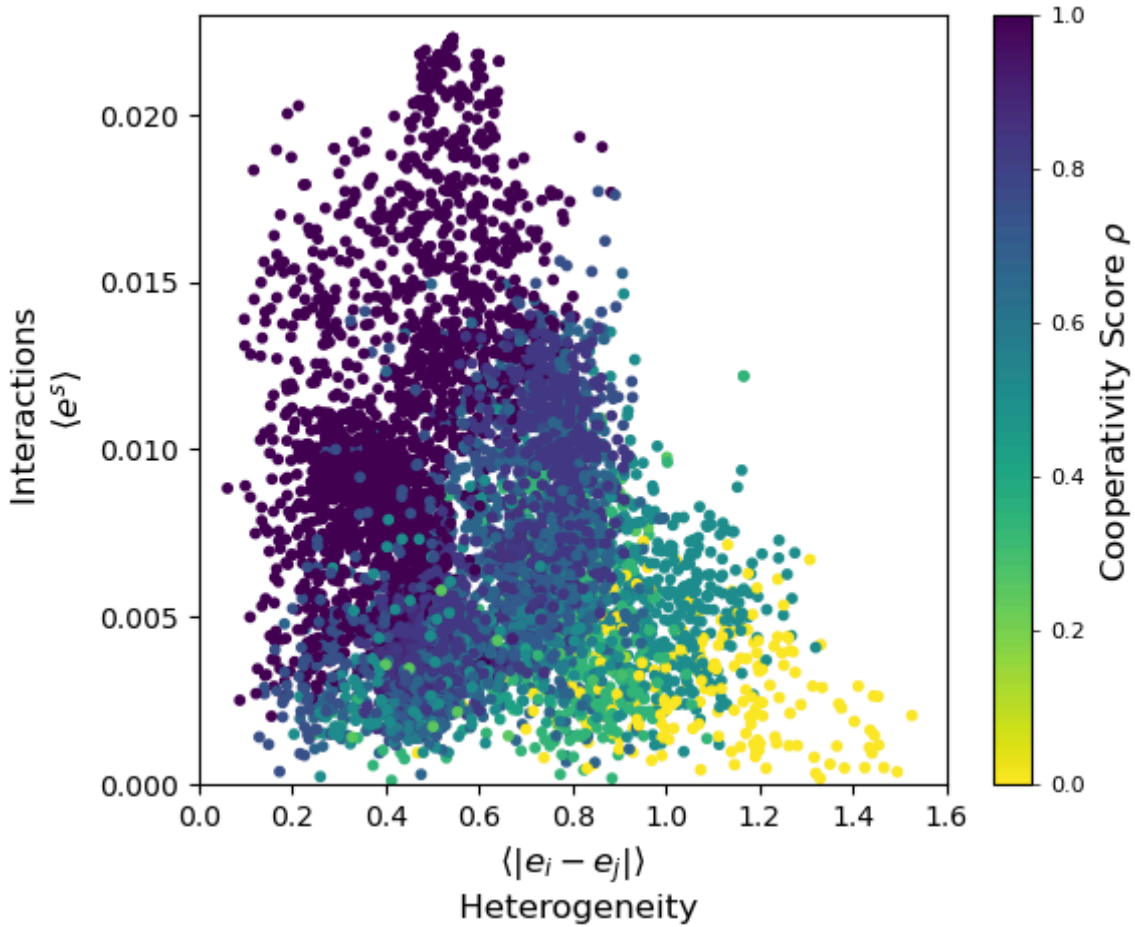

*Figure 3. Cooperativity Score for 7490 sequences of 15 different protein families, in a color scale on the space defined by the energetic heterogeneity of foldons and their average interaction strength.*

6. Extra: Exploit family results for protein design.

Once the Ising model has been set up for computing the folding dynamic of a given sequence, it is straightforward to also apply it for many other sequences of the same protein family. The evolutionary Potts model, the foldon assignment and Monte Carlo parameters are usually shared for the same family. Protein folding temperature and cooperativity are expected to vary according to the evolutionary energy assignment of the sequences [21]. However, depending on the tertiary structure topology some protein families can present very conserved mechanisms. By construction, sequences with a more favourable total energy will be more stable, so just computing and ranking the overall energy of sequence is enough to pick natural sequences that are expected to be more thermally stable, with potential biotechnological applications.

The cooperativity score is expected to decrease with foldon energetic heterogeneity and increase with the average interaction strength (figure 3). We have seen that using a linear fit, useful predictions can be made to estimate folding temperature and cooperativity changes upon single site mutations (figure 4). The folding elements assignment modulates the effect of local stability changes. For instance, a destabilizing mutation in a highly stable element reduces the energetic heterogeneity, producing a more cooperative folding (red vertical stripes in figure 4A). On top of that, the mutation impact on inter-element interactions also affects predictions. We have seen that the cooperativity variance of natural sequences within a protein family presents a positive correlation with the predicted cooperativity variance for point mutations [21]. Therefore, it is expected that $\alpha$-protein families are the most sensitive to engineer single variants with different folding cooperativity.

Moreover, the evolutionary Potts model that is being used to input the folding model can also be leveraged to generate an arbitrary amount of new protein sequences for the family using a Monte Carlo simulation [33]. All these generated synthetic sequences can be located in the heterogeneity-interactions phase space (figure 3) to predict their folding cooperativity even before computing a folding simulation, without any additional computational cost.

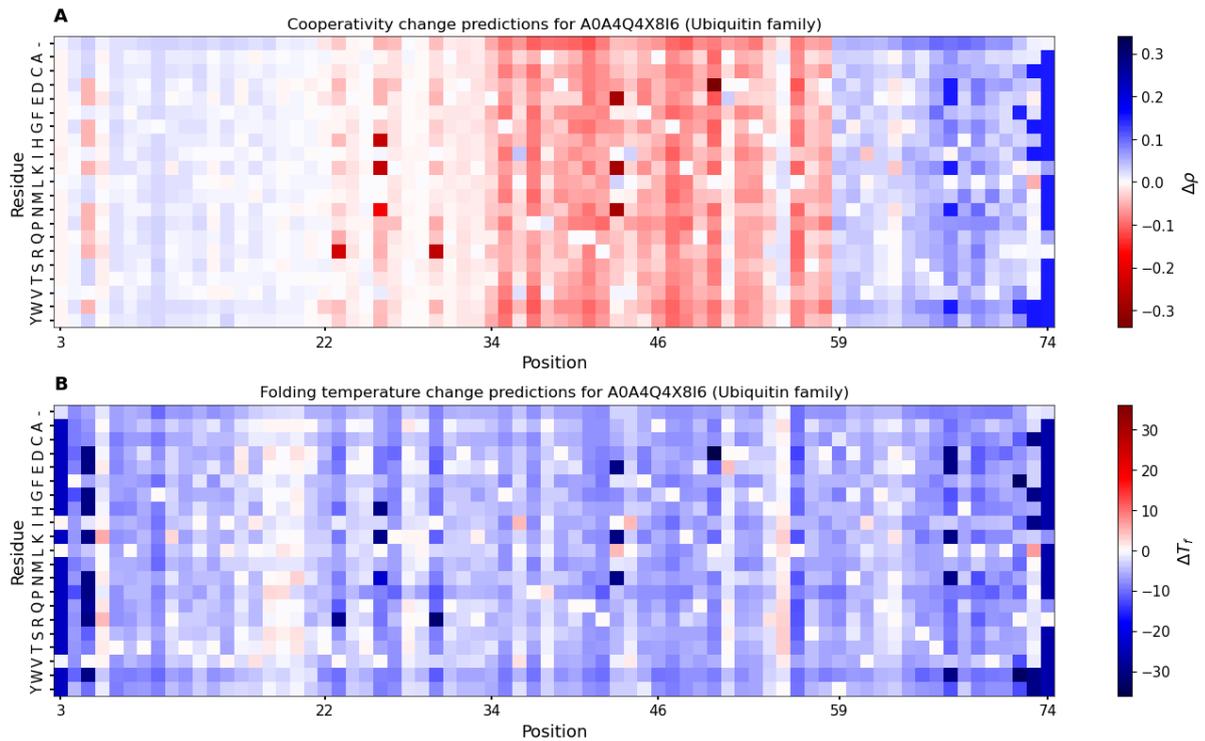

*Figure 4. Single site mutant predictions.* (A) Changes in the Cooperativity score upon single point mutations, for a natural sequence of the Ubiquitin family (Uniprot ID A0A4Q4X8I6). Foldon boundaries are indicated following reference PDB numbering (1ubq). (B) For the same sequence, the changes in the Folding Temperature upon all single-point mutations.

**Notes**

Note 1. We highlight that the proposed framework assumes that folding stability is locally the main evolutionary constraint, an approximation in line with the minimal frustration principle. Therefore, sequence positions strongly conserved and conditioned by other selection forces besides folding may affect local stability and some cooperativity predictions, locally frustrating the folding landscape [36].

Note 2. Alternatively, Potts model parameters (local fields and couplings) can be also obtained from other evolutionary sequence-based models, often faster to train or already trained and available. For instance, the weights of a Gaussian Restricted Boltzmann Machine (RBM) model can be exactly mapped to the Potts model parameters [37]. Furthermore, it has been tested that protein Large Language Models as ESM2 [38] 'logits' can be leveraged to calculate a categorical Jacobian, a useful approximation for Potts couplings [39].

Note 3. Folding Dynamics variability can emerge within protein family members or not according to the protein topology. As a rule of thumb, elongated alpha proteins allow different folding mechanisms depending on the sequence, while compact beta containing proteins do not. For the latter, a vanilla model that uses only contact positions and a binarized secondary structure (alpha/beta or coil/loop) may be enough to get the folding cooperativity, without requiring a sequence-sensitive Potts model [21].

**Acknowledgments**

This work was supported by the Consejo de Investigaciones Científicas y Técnicas (CONICET) (DUF is a CONICET researcher and EAG is a postdoctoral fellow); CONICET Grant PIP2022-2024—11220210100704CO and Universidad de Buenos Aires grant UBACyT 20020220200106BA. We call the attention of the international scientific community about the catastrophic erosion of Argentina's strong scientific tradition due to current funding constraints and the sudden termination of long term policies.